# Image Memorability Predicts Social Media Virality and Externally-Associated Commenting

Shikang Peng[1,2,3*], Wilma A. Bainbridge[3,4]

[1] Department of Psychology, University of Toronto, Toronto, ON, Canada

[2] Rotman Research Institute, Baycrest Health Sciences, Toronto, ON, Canada

[3] Department of Psychology, University of Chicago, Chicago, IL, USA

[4] Neuroscience Institute, University of Chicago, Chicago, IL, USA

*Corresponding author. Email: speng@research.baycrest.org

Shikang Peng https://orcid.org/0000-0002-0969-9642

Wilma A. Bainbridge https://orcid.org/0000-0002-7554-0736

## Abstract

Visual content on social media plays a key role in entertainment and information sharing, yet some images gain more engagement than others. We propose that image memorability – the ability to be remembered – may predict viral potential. Using 1,247 Reddit image posts across three timepoints, we assessed memorability with neural network ResMem and correlated the predicted memorability scores with virality metrics. Memorable images were consistently associated with more comments, even after controlling for image categories with ResNet-152. Semantic analysis revealed that memorable images relate to more neutral-affect comments, suggesting a distinct pathway to virality from emotional contents. Additionally, visual consistency analysis showed that memorable posts inspired diverse, externally-associated comments. By analyzing ResMem's layers, we found semantic distinctiveness was key to both memorability and virality. This study highlights memorability as a unique correlate of social media virality, offering insights into how visual features and human cognitive behavioral interactions are associated with online engagement.

**Keywords**: memorability, vision, virality, social media, convolutional neural network, natural language processing

# 1. Introduction

What makes something go viral on social media? When browsing various social media platforms (e.g., Reddit, X/Twitter, Instagram), you may have seen some posts receive more upvotes or comments than others. Prior research has shown that positive, intense emotion, or inclusion of moral messages in a post may lead to virality, as revealed by the number of upvotes and comments (Alhabash et al., 2015, 2019; Berger & Milkman, 2013; Heimbach & Hinz, 2016). These elements appear to be associated with increased cognitive engagement, as evidenced by prioritized visual attention and unique physiological response patterns to viral posts (Alhabash et al., 2015; Brady et al., 2020). This suggests that the ability of content to go viral may depend on its impact on cognitive processes. Although the above studies have found emotion and morality to be significant factors, these factors stem from a subjective appraisal of the post and thus may not exert a permanent and consistent influence on viral behavior across people. For instance, provocative content might momentarily capture attention, or a morally charged post may go viral among individuals who are highly engaged with current events of sensitive to polarizing topics, or a specific culture group but not others. Such subjective or context-dependent elements, therefore, may not reliably or consistently predict media virality across diverse audiences. Also, while emotional content has been linked to virality, it is likely not the only contributing factor. Identifying additional predictors that account for more variance and that generalize more reliably across people, is an important direction. In search of a more consistent, human-originated cognitive predictor, we propose that visual properties intrinsic to the image itself – features inherent to the image's content and composition, independent of external factors like context or culture – may offer a more stable and universally applicable predictor of virality. Specifically, we focus on image memorability, or how effectively an image sticks in human memory.

Images possess an intrinsic memorability, defined as the likelihood that an image will be remembered by a consistent proportion of people (e.g., remembered by 80% vs. 20% of viewers). In other words, people consistently remember high memorability (memorable) images and forget low memorability (forgettable) images (Isola et al., 2011b; Bainbridge et al., 2013). This effect has been replicated across several image types, including scenes (Isola et al., 2011b), faces (Bainbridge et al., 2013), objects (Kramer et al., 2023) and artwork (Davis & Bainbridge, 2023). Different from simple low-level visual features like color or salience, memorability is found to be a unique high-level visual feature separate from other high-level features like emotion (Bainbridge et al., 2013, 2017; Bainbridge & Rissman, 2018). As such, memorability could in theory capture a separate aspect of virality from what has been previously identified with emotion and morality. The memorability of an image has also been shown to be remarkably consistent across cultures (Jeong, 2023), age groups (Guo & Bainbridge, 2023), tasks (Bainbridge, 2020), and environments (Davis & Bainbridge, 2023), suggesting that it is very stable across people and contexts. Additionally, research has indicated that memorable items are more likely to be remembered even when participants are not paying attention (Wakeland-Hart et al., 2022; Roberts et al., 2024) or not performing a memory task (Bainbridge, 2020), implying that memorability can affect human perception automatically. This is consistent with the idea that viral behaviors may stem from unconscious, automatic responses, such as pressing the "like" button, driven by genuine interest rather than a deliberate intention to increase virality (Alhabash et al., 2019). Hence, we hypothesized a positive correlation between an item's memorability and its likelihood of going viral, as more memorable items are more likely to automatically stick in memory and gain widespread attention.

Leveraging a state-of-art model named ResMem, a specialized convolutional neural network (CNN) trained with human data to predict an image's memorability score (Needell & Bainbridge, 2022), here we investigated the predictive relationship between model-predicted image memorability and social media virality. This approach is justified by previous evidence that ResMem's memorability prediction effectively predicts human memory performance across diverse domains, including artwork, real-world objects, and scenes (Kramer et al., 2023; Davis & Bainbridge, 2023; Koch et al., 2020; Zhao et al., 2024), suggesting its predictions could also generalize to social media posts and audiences. Within the above findings, one notable finding reveals that even without training on artwork or cultural knowledge, ResMem predicted higher memorability scores for famous artworks versus non-famous artworks in the Art Institute of Chicago's collection (Davis et al., 2023). This supports our hypothesis that part of what makes an image famous and potentially viral, is its ability to last in memory. Further, examining the representations within the convolutional layers of such a CNN could promote a richer understanding of how different types of image features contribute to the final memorability estimate. As prior research has shown early convolutional layers are associated with low-level visual features and later layers with high-level semantic features (Yamins & DiCarlo, 2016; Kriegeskorte, 2015; Khosla et al., 2015; Xu & Vaziri-Pashkam, 2021), assessing these layers' representations may reveal whether the factors driving memorability are also key indicators of viral behaviors. This approach thus offers a more precise understanding of the mechanistic link between memorability and viral success.

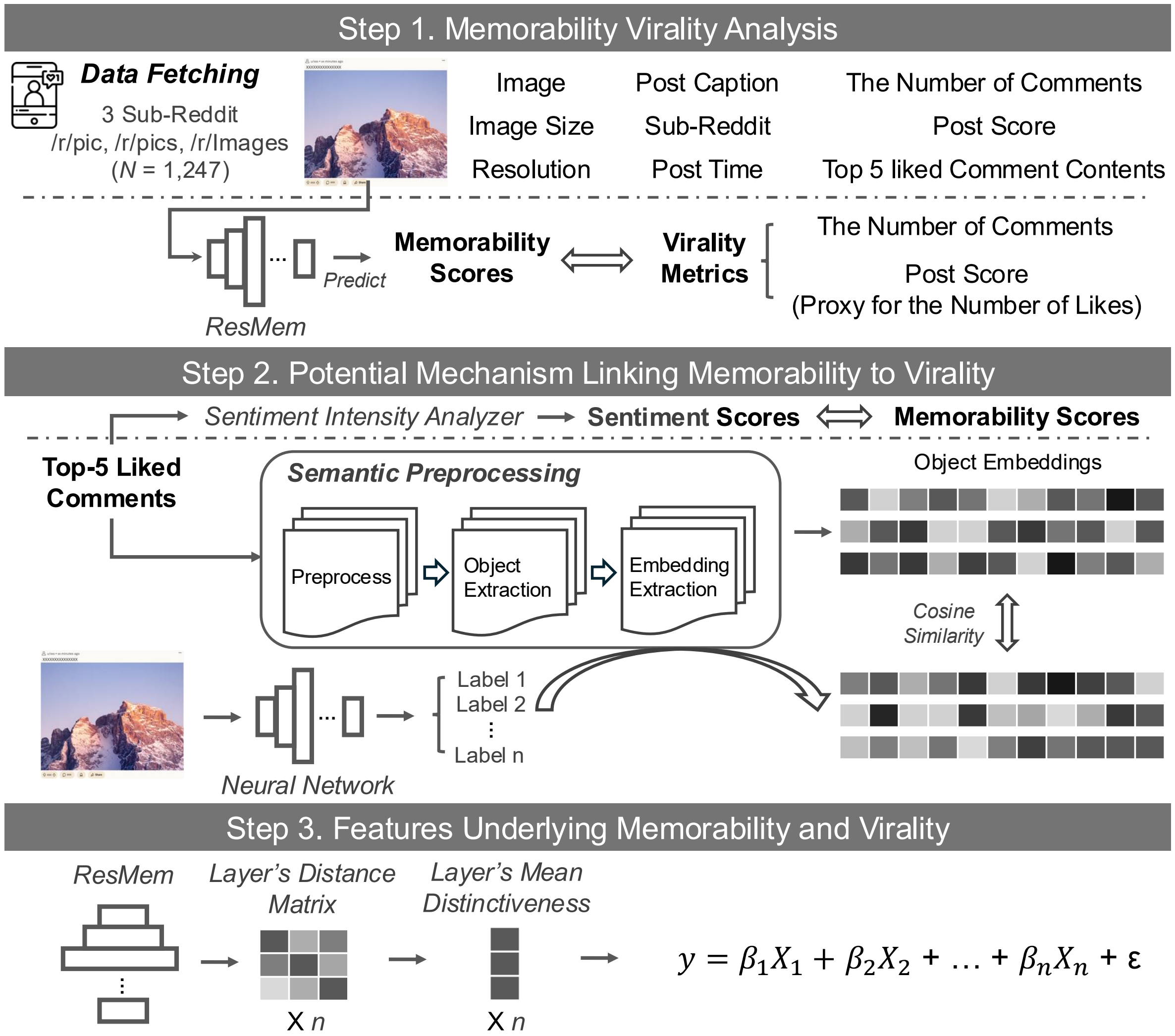

**Fig 1. Diagram of the overall study workflow.** Example of a social media post resembling those from any of the three selected image-based subreddits (communities) on Reddit with upvote/downvote (like/dislike) and comment buttons, with metadata of each post. The image shown in the figure is for explanatory purposes only. Each post is associated with nine variables including: 1) image; 2) image size; 3) post caption; 4) caption length; 5) sub-reddit; 6) post time; 7) the number of comments; 8) post score (a proxy of the number of upvotes); 9) the contents of the top five liked comments. The Step 1 analysis tests how image memorability relates to measures of virality, specifically the number of comments a post receives and its post score. The Step 2 analysis examines the relationship between memorability and the sentiment and semantic content present within the top-5 comments of each post. The Step 3 analysis tests how semantic distinctiveness within the ResMem CNN relates to memorability and virality.

The present study centers around the hypothesis that image memorability predicts media virality, specifically addressing two main research questions: 1) How does image memorability predict virality?; and 2) What underlying image features contribute to this predictive relationship? (**Fig. 1**) To explore these questions, we collected a naturalistic dataset of 1,247 Reddit image posts sampled across three distinct timepoints from three image-based communities, allowing us to investigate image memorability effects in real-world social media contexts. We leveraged the ResMem model, trained to predict image memorability, to estimate memorability scores for each image, a novel approach in the context of studying social media virality in the real-world. Correlational analyses revealed that memorable images consistently attracted more comments, a finding replicated across timepoints. Crucially, even after controlling for image category effects using ImageNet pre-trained ResNet-152, ResMem's memorability-specific components still explained unique variance in virality. We then conducted semantic analyses of the post comments to understand *how* memorability predicts virality. Results revealed that memorable images were linked to more neutral-affect, externally-associated (information that is not directly presented in the image), and diverse discussions (e.g., a dragon sculpture may evoke a watermelon or a phobia; **Fig. 2AB**), suggesting that memorability may engage viewers differently than emotional valence. Further analysis of image representations in ResMem's convolutional layers indicated that semantic distinctiveness was associated with both memorability and virality, pointing to a shared underlying feature. In sum, this study highlights the relationship between semantic distinctiveness, memorability, and virality, revealing a novel association between intrinsic image features and social media interactions.

Overall, this work contributes four key advances. First, we introduced a novel application of image memorability prediction to the study of real-world social media virality. Second, we provided evidence that memorability explains unique variance in virality beyond image category effects. Third, we presented a semantic analysis showing that memorability predicts externally-associated and diverse online discussions, a pattern not captured by conventional emotion-based models. Finally, we offered insights into semantic distinctiveness as a potential mechanism linking memorability and virality.

## 2. Methods

**2.1. Materials.** Data was fetched from the popular English social media platform Reddit (over 500 million users as of 2024), where users can post images, videos, text, or links and react to posts by giving "upvotes/downvotes" or commenting below the post. The collection process was automatically completed by a customized Python scraper script. The data collection targeted acquiring 600 images across three time points, with the first on Apr 19, 2024, the second on Apr 26, 2024, and the third on May 21, 2024. Data from three time points was collected to ensure the replicability of any effects across different attempts. We applied customized functions in the scraper to avoid collecting duplicated posts and unqualified posts based on the criteria in the next paragraph. Overall, 1,247 valid image posts were fetched from Reddit through random sampling, with 504 images for the first collection, 367 images for the second, and 376 for the third. This sample size was chosen based on prior findings suggesting that machine learning models for predicting image virality tend to reach stable performance after training on around 1,000 images, with minimal gains beyond that point using basic image features (Deza & Parikh, 2015).

Posts were selected based on the following criteria: 1) at least five upvotes and five comments to exclude non-informative posts for subsequent analysis; and 2) exactly one image was contained in the post. The data was fetched equally from three general-purpose image subreddits: */r/pics, /r/pic, /r/images*, to enhance the generalizability of any effects. These subreddits are communities in Reddit that require all posts to have one image and tend to feature images on a variety of topics, such as a photo of land with a title like "Crown land camping. Ontario" (**Fig. 2A**). All three subreddits used in this study are in at least the top 2% of communities ranked by size, with */r/pics* being in the top 12 subreddits across the whole platform. We collected the following data for each post that we fetched: 1) image; 2) image size; 3) post caption; 4) caption length; 5) sub-reddit; 6) posting time; 7) the number of comments; 8) post score (a proxy of the number of upvotes); 9) the contents of the top five liked comments.

**2.2. Virality Analysis.** To examine if higher memorability predicts higher engagement, each image in the dataset was processed through the pre-trained ResMem model (version 1.1.6, https://pypi.org/project/resmem/) to obtain a predicted memorability score (**Fig. 1**). High-memorability images were defined as those with scores above the sample median, while low-memorability images fell below the median. We then computed the Spearman rank correlation to assess the relationship between memorability scores and the two engagement metrics: the number of comments and upvotes each image post received. Spearman correlation is a non-parametric statistics that evaluates the strength and direction of association between two variables based on their ranked values, making it appropriate for non-normally distributed data such as social media engagement. While rank correlation is robust to outliers, we were concerned that extreme values might introduce overdispersion in the subsequent regression models used for analyses of neural layer contributions and in the semantic assessments of the comments. These extreme values, though informative about engagement levels, could still potentially skew the results, which has been specifically noted in social-media related research (Heimbach & Hinz, 2016). To address this concern, we conducted two assessments of the data: 1) without outlier removal: we analyzed the entire dataset as it was, with a total of 1,247 images; 2) with outlier removal: We cleaned the data by removing outliers where either the number of upvotes or comments in the post were more than 1.5 times the interquartile range (IQR) above the upper quartile or below the lower quartile, resulting in a total of 997 images. Furthermore, we conducted the same analysis within each subreddit to ensure the found effect is not only limited to one community.

We calculated and averaged the sentiment scores of the comments in each image post through the pre-trained sentiment intensity analyzer from Python package Natural Language Toolkit (Bird et al, 2009). This toolkit analyzes a given text and provides a dictionary of sentiment scores, where the compound score, ranging from -1 (most negative) to 1 (most positive), represents the overall sentiment of the text. This analysis aimed to investigate whether image memorability may predict the subsequent emotional responses expressed in the comments.

To examine whether potential confounds might influence the relationship between memorability scores and virality, we computed a correlation heatmap including several media post covariates: number of comments, averaged sentiment score, caption length (number of letters in each post title), time of day (a dummy-coded variable where day = 0 and night = 1), posted duration (the number of days since the post was published online), file size (in kilobytes), and image resolution (calculated as the product of width and height in pixels), along with predicted memorability scores and engagement metrics (**Fig. 4C**). Covariates found to be significantly correlated with memorability and engagement were then controlled for in a partial correlation analysis between memorability and virality.

**2.3. Consistency between image content and comments**. To obtain a measure of the image-comments relationship, we pursued a fully automated approach through the use of Google Cloud Vision API (Google, n.d.), a visual CNN. The label detection function of this API takes in an image and gives an output of word labels representing the entities in the image. These labels were then tokenized and lemmatized, and only the unique words within a list were preserved.

Comment data was first tokenized and lemmatized to obtain segmented text data. We focused exclusively on nouns (**Fig. 2AB**), as these are the part of speech that can be directly compared with the object labels to provide comparable measures (Misra et al., 2016). Subsequent preprocessing involved the use of Python package *spaCy's* (Honnibal & Motani, 2017) English words dictionary with the trained pipeline "en_core_web_sm" to remove non-English words and uninformative words from a custom list of stop words. These custom stop words included words found to be the synonyms of "image" or acronyms that are not recognized properly by the package, identified through manual screening (e.g., pic, picture, post, lol) (**S2,6**). Additional preprocessing was taken to further remove non-semantic text features, including emojis, URLs, symbols, and to keep only one instance of the words that repeated three or more times consecutively (i.e., "Money Money Money" will result in "Money"). This ensured that only the core and informative semantic features were preserved for each image post.

To compare the lists of image labels and comment tokens created from the last section, we used word embeddings as they capture the semantic relationship between them even when their physical forms differ (e.g., "tower" vs. "building"), ensuring the comparability of these two lists of words (**Fig. 2B**).

To start, we leveraged the package named Global Vectors for Word Representation (GloVe) (Pennington et al., 2014). This is an unsupervised learning algorithm designed to obtain vector representations for words. Specifically, we applied a pre-trained GloVe word vector model (6B.100d, https://nlp.stanford.edu/projects/glove/) that was trained on 6 billion tokens, with a vocabulary of 400,000 words, each represented in a 100-dimensional space. This pre-trained vector takes a word as an input and returns its 100-dimension vector representation. To compare the embeddings, we calculated the cosine similarity in the vector space between each pair of image label and comment word. The highest correlation between a label in the image and a word in the comment indicates the closest match between the image content and comment. For instance, if a "rock" appears in the comment, it will get closely matched to "stone" in the image rather than "ground" even though they do share some semantic similarity. This algorithm guarantees the precision of the comparability between image labels and comments. After that, this correlation was averaged across all the tokenized comments' highest match cosine similarity to represent the consistency level between the images and their comments.

**A**

**High Memorability Image**

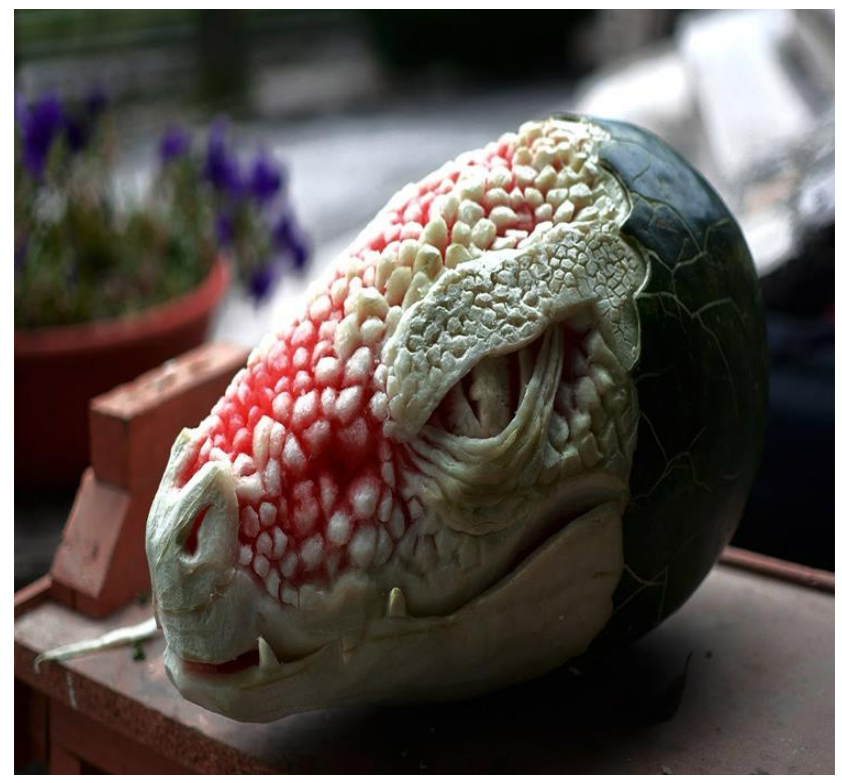

**Low Memorability Image**

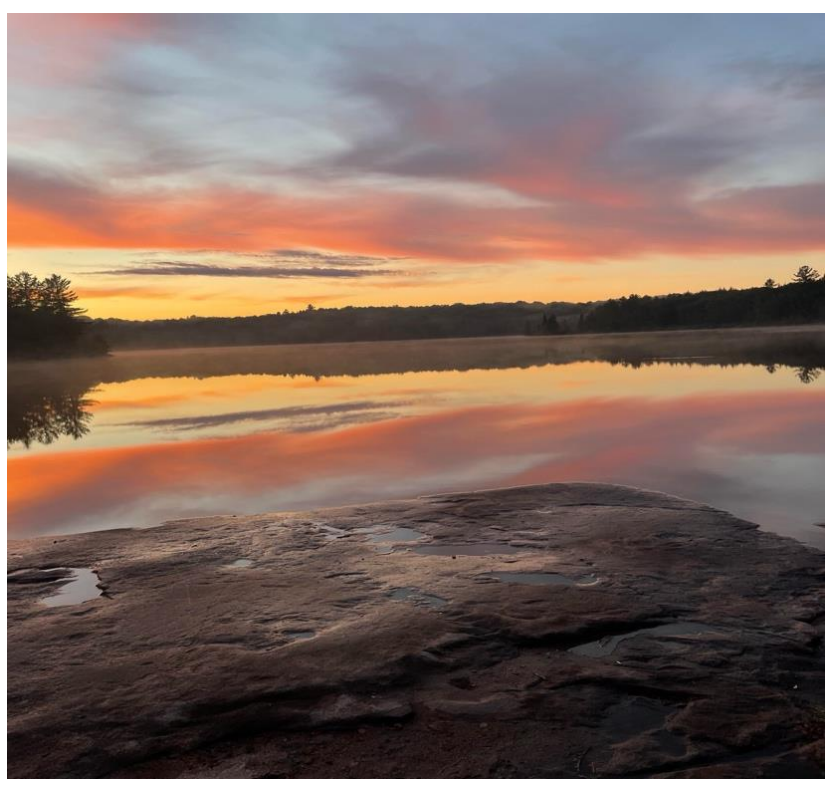

“'It looks like a very angry **watermelon**'”
“'Oh no, this **watermelon** was touched by D I E G O B R A N D O”
“'Damn this is amazingly cool. Super **talent** here”
“'**Water dragon**!”
“This is amazing! but it gives me that **t-word phobia** feeling under my **skin**. I can't remember the **word** and I am certainly not googling it."

“**Mosquitos**?”
“**Trout**”
“I prefer ‘public **land**‘,great pic”

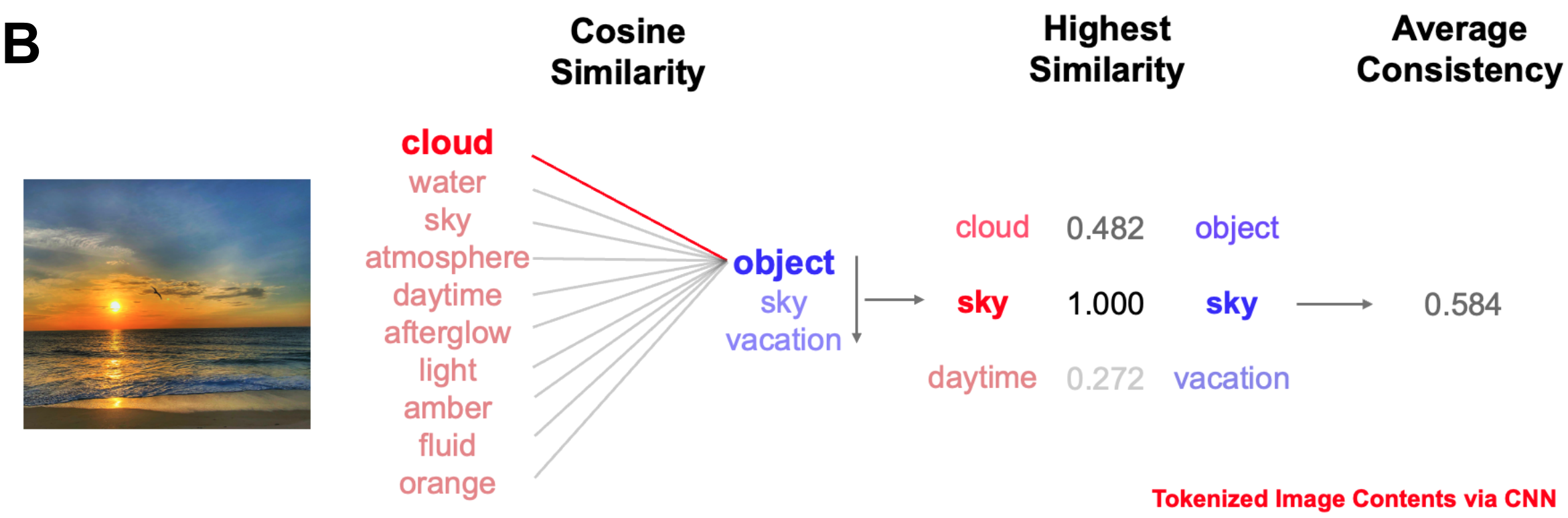

**Fig. 2. Image Contents and Comments. (A)** The top 5 comments with nouns (highlighted) associated with a high memorability and low memorability image. **(B)** To quantify the consistency level between the image contents and comments, each tokenized word in the comments was iteratively compared with the tokenized image contents in the CNN-recognition list in their vector space through cosine similarity. When a highest similarity was found, indicating a closest match between a word in the comments and an object in the image, it was stored and then averaged across other highest cosine similarity from other words in the comments.

**2.4. Convolutional Layer Selection.** To quantify the contribution of underlying visual and semantic representations towards the memorability estimates and social media metrics, we first focused on the architecture of the CNN. ResMem is a CNN that merges and concatenates the architectures of ResNet-152 and AlexNet, passing them through 3 fully connected layers to generate a memorability score estimation (**Fig. 3A**). ResNet-152 is known for its deep residual learning framework, allowing it to effectively handle more complex underlying features (He et al., 2015), while AlexNet is one of the pioneering CNN architectures that demonstrated the power of deep learning in visual tasks (Krizhevsky et al., 2012). The architecture of the above two CNNs are widely used in the visual recognition field, and they are found to resemble the processing steps of the human visual processing stream (Yamins & DiCarlo, 2016; Kriegeskorte, 2015; Khosla et al., 2015; Xu & Vaziri-Pashkam, 2021).

Specifically, the components within the models were retrained for predicting image memorability, allowing ResMem to adapt and refine the originally pre-trained ResNet-152 and AlexNet features. As such, these features, initially developed for image category classification with ImageNet (Deng et al., 2009) (a large-scale visual database designed for use in visual object recognition research), are now re-optimized to capture determining features of image memorability. Given that, this investigation extracted layer output straight from ResMem to allow for inference about the features contribution to memorability. Based on the visualizations of features in the most informative late convolutional layers of the AlexNet component of ResMem (Needell & Bainbridge, 2022), weights in the shallower AlexNet are primarily driven by low-level features like orientation and color. In contrast, the feature visualizations of the ResNet-152 display a wide variety of patterns and textures, reflecting the network's ability to capture more complex and diverse semantic content due to its depth. The early layers of ResMem primarily focus on basic patterns and textures similar to those in AlexNet, but as we progress to the late layers, the geometrical patterns become increasingly intricate, often resembling objects (e.g., eyes, arches). As such, to gain a richer understanding of how memorability is perceived by ResMem, the analyses in this manuscript only focused on the convolutional layers of the ResNet-152 branch as it provides a more detailed and comprehensive examination of semantic information processing beyond low-level visual features.

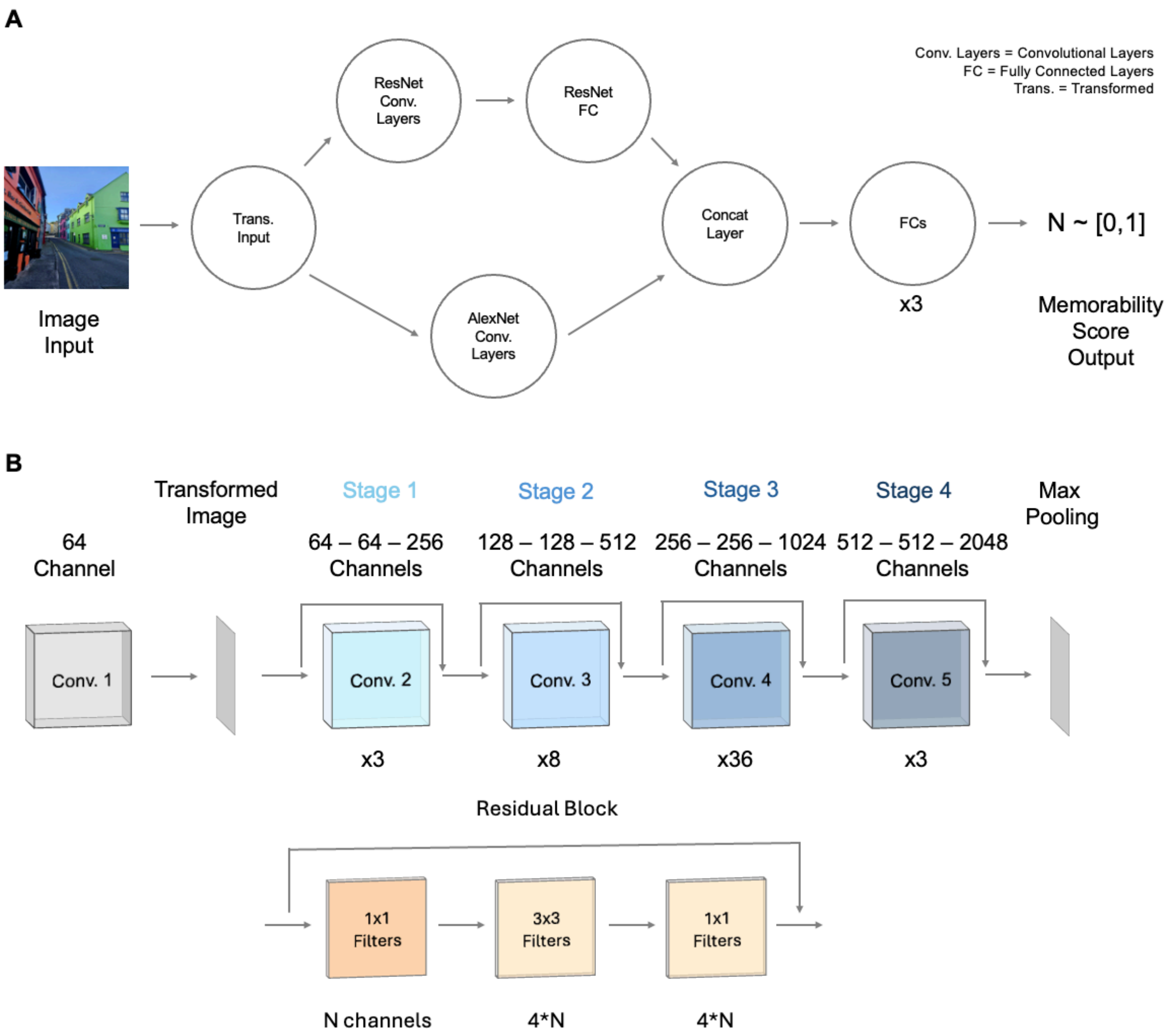

**Fig. 3. Convolutional neural network architecture. (A)** Architecture of ResMem. Image input is processed by both ResNet-152 and AlexNet networks. The output of the two CNNs are concatenated together and processed through three fully connected layers to generate the final memorability estimate output. **(B)** Architecture of ResNet-152. The image input first goes through a convolutional layer for initial processing. The filtered image then goes through four stages of residual blocks which contain three convolutional layers in each. The output of the convolutional stages is processed by a max pooling layer to reduce the spatial dimensions of the input feature maps while retaining the most significant information.

To unveil how ResMem perceives memorability using the ResNet-152 architecture, we structured the feature-extraction process in four key stages. This approach mirrors prior studies using shallower CNNs (Xu & Vaziri-Pashkam, 2021; Koch et al., 2022; Kramer et al., 2023), with each stage corresponding to the four major residual blocks in ResNet-152 (**Fig. 3B**): 1) low-level visual features extraction; 2) feature integration; 3) abstraction; and 4) high-level conceptual synthesis. The naming is based on feature visualizations of different convolutional layers, reflecting the well-established progression from low-level visual features to more abstract representations (Needell & Bainbridge, 2022; Yamins & DiCarlo, 2016; Kriegeskorte, 2015; Khosla et al., 2015; Xu & Vaziri-Pashkam, 2021). In each stage, to identify the most representative convolutional layer, it is essential to examine the arrangement and contributions of residual blocks first. Although the dimensionality of the convolutional layers is amplified in each residual block and then normalized to reduce the dimensionality to match the input dimension of the next block, crucial information is progressively refined and carried forward. Hence, we considered the last convolutional layer of the last residual block in each stage as the most representative layer due to its role in finalizing the feature transformations of that stage.

Given the third stage of ResNet-152 comprises 36 residual blocks, which is significantly larger than the others, we further divided this stage into three evenly distributed sub-stages for more nuanced analysis: 1) early: abstraction initiation; 2) middle: intermediate abstraction; 3) late: abstraction refinement. We selected the last convolutional layer in each one-third point of the residual block to capture the most representative features at different phases during the middle-level abstraction processes. In total, six neural layers were selected for subsequent analysis.

**2.5. Quantifying the contribution of convolutional layers to memorability and social media metrics.** To assess the contribution of each network learning stage to the final memorability output, we began by extracting the outputs from each representational convolutional layer for all images in the dataset through ResMem. A Pearson correlation was applied to the flattened vectors from these convolutional layer outputs to generate a distance matrix, in which each row and column indicates the dissimilarity (1 – Pearson r) between one image and another (**Fig. 5AB**). We then averaged the dissimilarity scores across each row (excluding on-diagonal cells to exclude correlations with oneself) to obtain a mean dissimilarity score for each image (**Fig. 5C**). This score quantifies the relative distinctiveness of each image compared to others in the dataset. To avoid the confounding effects of image category on memorability estimates, we utilized the original ImageNet-pretrained ResNet152. Following the same approach applied to ResMem, we regressed out the category-specific components of ResNet from ResMem through ordinary least square regression (OLS). Specifically, for each ResMem layer, we performed a univariate OLS regression using the corresponding ResNet layer as the predictor to isolate the variance in ResMem not explained by its corresponding ResNet feature. This process resulted in a ResMem-residual, which represents the component contributing solely to memorability. These residuals were then fitted into a generalized linear gaussian regression model (GLM) with the dependent variable as the image's memorability score to analyze how the extracted features from each learning stage influence image memorability. We tested for multicollinearity in the mean dissimilarity scores across the different neural layers through computing the variance inflation factor (VIF) and identified no strong evidence for multicollinearity (all below 5).

Prior work has found that convolutional layers tend to represent different levels of visual information, with early layers showing more low-level details and late layers representing more high-level abstract and semantic information (Yamins & DiCarlo, 2016; Kriegeskorte, 2015; Khosla et al., 2015; Xu & Vaziri-Pashkam, 2021). As such, by fitting the mean dissimilarities for images per layer into a GLM, we may gain a more nuanced understanding of how different levels of visual information may contribute to subsequent social engagement metrices. A GLM with a negative binomial distribution was fitted to the number of post scores and comments to further ensure no overdispersion issue raised by the heavily skewed count nature of social media data (Heimbach & Hinz, 2016; McElreath, 2020), while a Gaussian distribution was fitted to the averaged sentiment scores.

## 3. Results

**3.1. Memorability is associated with viral social media posts**. To examine the relationship between memorability and virality, we analyzed 1,247 posts from three general-purpose image-based Reddit communities, collected across three distinct timepoints to support replication (**S1**). For each post, we extracted the image, post score, the number of upvotes and comments, and the top five most liked comments. Notably, the post score was used as a proxy for the number of upvotes (how many people liked the post), with further details provided in the Discussion section .

There was a significant positive Spearman rank correlation between memorability and the number of comments for all three timepoints ($\rho_1$=.183, p<.001; $\rho_2$=.135, p=.009; $\rho_3$=.329, p<.001). These results replicated even when we used 1.5 IQR to remove outlier posts from our dataset ($\rho_1$=.157, p=.014; $\rho_2$=.132, p=.025; $\rho_3$=.234, p<.001). Similar results were found when we conducted correlations within each subreddit (without outlier removal: $\rho_{pic}$=.307, p=<.001;

$\rho_{pics}=.128$, $p=.002$; $\rho_{images}=.137$, $p=.008$; with outlier removal: $\rho_{pic}=.261$, $p=<.001$; $\rho_{pics}=.132$, $p=.003$) with one smaller effect that did not reach significance (with outlier removal: $\rho_{images}=.044$, $p_{images}=.425$). Overall, the positive relationship between post memorability and number of comments remained consistent. We also tested the relationship between memorability and post score and observed a less robust relationship. Without outlier removal, memorability had a positive association with post scores in two out of three timepoints ($\rho_1=.103$, $p=.021$; $\rho_2=.133$, $p=.001$; $\rho_3=.030$, $p=.56$). With outlier removal, the correlations were mixed across the three timepoints ($\rho_1=.059$, $p=.233$; $\rho_2=.138$, $p=.019$; $\rho_3=-.193$, $p=.008$). Similar unstable results were found when we conducted correlation within each subreddit (without outlier removal: $\rho_{pic}=.091$, $p=<.001$; $\rho_{pics}=.128$, $p=.002$; $\rho_{images}=.036$, $p=.486$; with outlier removal: $\rho_{pic}=-.078$, $p=.042$; $\rho_{pics}=.047$, $p=.299$; $\rho_{images}=-.034$, $p=.532$).

Overall, the results show a robust positive correlation between memorability and the number of comments, but not the post scores. To simplify further analysis, we concatenated data across all timepoints and removed the outliers. The significant positive correlation between memorability and number of comments persisted ($\rho=.203$, $p<.001$) (**Fig. 4A**).

As shown from the correlation heatmap **(Fig. 4C)**, only caption length and resolution showed small significant correlations with both memorability and the number of comments. However, a partial correlation analysis between memorability and the number of comments controlling for caption length and resolution still showed a significant correlation ($\rho=.120$, $p<.001$).

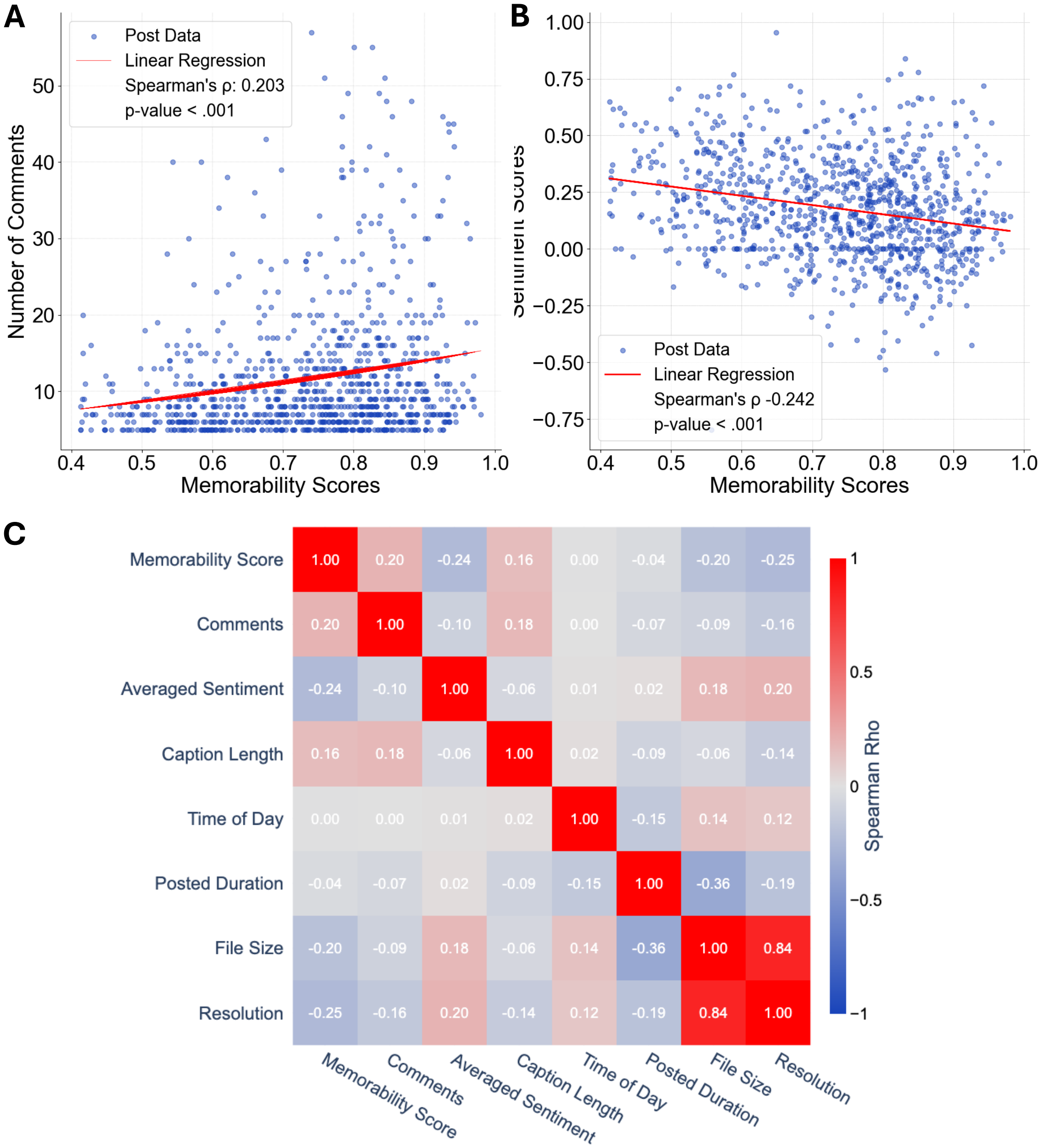

**Fig 4. Media Engagement Prediction and Memorability Covariates. (A)** Concatenated post data across three timepoints with outliers removed, with a total of 997 images. Image memorability had a significant positive association with the number of comments. **(B)** Memorability of image posts showed a significant negative correlation with the average sentiment scores of the comments. (**C**) Spearman correlation heatmap of covariates with memorability scores. Averaged sentiment is defined as the average sentiment score computed by sentiment intensity analyzer for each image post. Caption length is measured as the number of letters in the post title. Time of day is a dummy-coded variable (Day = 0, Night = 1). Posted duration is the number of days since the post was published online. File size is measured in kilobytes (KB). Resolution is defined as the product of width and height in pixels.

**3.2. Memorable posts lead to neutral and externally-related comments.** To explore the relationship between memorability and virality, we examined how memorability is associated with the nature of discussions about the images. Sentiment analysis of the comments revealed a significant negative rank correlation between image memorability and sentiment valence ($\rho$ =-.242, $p<.001$). This relationship remained robust even after adjusting for significant covariates such as resolution and file size ($\rho$=-.237, $p<.001$; **Fig 4C**). Memorable images had a lower average sentiment score, closer to neutral valence and low intensity (0), compared to forgettable images (memorable = 0.134, forgettable = 0.222). This, combined with a significant negative association between memorability and sentiment intensity (the absolute value of sentiment scores; $\rho$=-.189, $p<.001$), highlights that while memorable images are associated with higher engagement, such engagement is linked to comments that are less positive or more emotionally neutral rather than negative attitudes (**Fig. 4B**). This finding demonstrates that memorability relates to virality in a different pathway from emotion (Alhabash et al., 2015, 2019; Berger & Milkman, 2013).

To assess the alignment between image content and comments, we computed the consistency score using cosine similarity between the vector representations of comment words and image object labels. This score ranges from 0 (no overlap between comments and image content) to 1 (full alignment between comments and image content). Our analysis revealed a significant negative correlation between image-comment consistency and memorability ($\rho$=-.116, $p$=.006), suggesting that the more memorable an image is, the more divergence there is between the image contents and comments. While longer comments may be more likely to diverge from an image's content, our analysis revealed the same negative relationship between memorability and image-comment consistency after accounting for comment length ($\rho$=-.127, p=.002). Along with the above analysis, this implies that higher memorability images tend to have comments focusing on external associations outside the image while lower memorability images are linked with comments focusing on the contents within the image.

**3.3. The primacy of semantic distinctiveness on memorability and engagement metrices.** To gain deeper insights into the mechanism by which memorability relates to virality, we analyzed the representations of these images within ResMem. As ResMem, a CNN derived from ImageNet pre-trained ResNet-152, is trained explicitly to estimate image memorability, its convolutional layers' outputs should inherently contain predictive information about memorability. Prior work has debated whether the distinctiveness or prototypicality (shared similarity across items) of visual features (physical characteristics like color or shape) and semantic features (abstracted characteristics related to meaning, category, and function) contributes to memorability, using simpler CNNs not specifically tailored for memorability (Koch et al., 2020; Kramer et al., 2023). However, ResMem, given its specificity in predicting memorability, provides direct and more accurate insights into the characteristics that make an image memorable. This analysis thus allows us to infer how these image representations within ResMem may affect virality, offering a more precise understanding of the relationship between memorability and viral potential.

However, image category may be partially confounded with the effect of memorability as certain image categories may be more likely to be memorable and/or viral. To avoid this possibility, we accounted for the image category effect by regressing ResMem-derived distances with ImageNet pre-trained ResNet correlation distances. As expected, results did not change meaningfully when using the ResMem outputs or the residuals regressing out ResNet outputs. Moreover, the residuals still explained unique variance, as indicated by their $R^2$ values (**Fig. 5D**), suggesting that ResMem's predictive abilities are not primarily driven by image category. Since these residuals isolate variance that is independent of image category, they are assumed to capture memorability-specific components. Thus, only the resulting residuals will be interpreted in the following analyses.

To unveil why memorability is related to human behavior on social media, we utilized this distinctive representation to bridge the gap between memorability-related features in the image and the engagement metrics (**S3-5**). Running a similar regression to predict the number of comments, we found that the more semantically distinctive an image is, the more comments it received ($\beta_{stage\ 3\text{-}early}$=3.213, p<.001, 95% CI =.164-4.784; $\beta_{stage\ 3\text{-}middle}$=-2.515, p<.001, 95% CI =-3.748 to -1.282; $\beta_{stage\ 4}$=1.062, p=.011, 95% CI =.247-1.807; other stages, p>.05; **Fig. 5E**). Though it is also clear that for low-to-middle level information, an opposite contribution to the number of comments was observed, which will be elaborated in the Discussion section. As expected, none of the convolutional layer representations were predictive of the number of upvotes. To examine how distinctiveness of semantic image features is related to commenting behavior, we also analyzed the sentiment in the comments. Similar to the correlation with the number of comments, the latest neural layer found a significant yet negative prediction with sentiment ($\beta_{stage\ 4}$=-0.395, p=.003, 95% CI =-.654 to -.135; **Fig. 5F**), suggesting that the semantic distinctiveness of images may provoke lower sentiment valence expressed in the comments. Meanwhile, distinctiveness within the middle stage that processes middle-level visual and semantic information positively contributes to sentiment ($\beta_{stage\ 3\text{-}early}$=-.625, p=.018, 95% CI=-1.142 to -0.107; $\beta_{stage\ 3\text{-}middle}$=0.675, p=.001, 95% CI=0.259-1.09). These results suggest that whereas more distinctive semantic representations result in more memorable images and more comments, visual and semantic features may have mixed impact on sentiment.

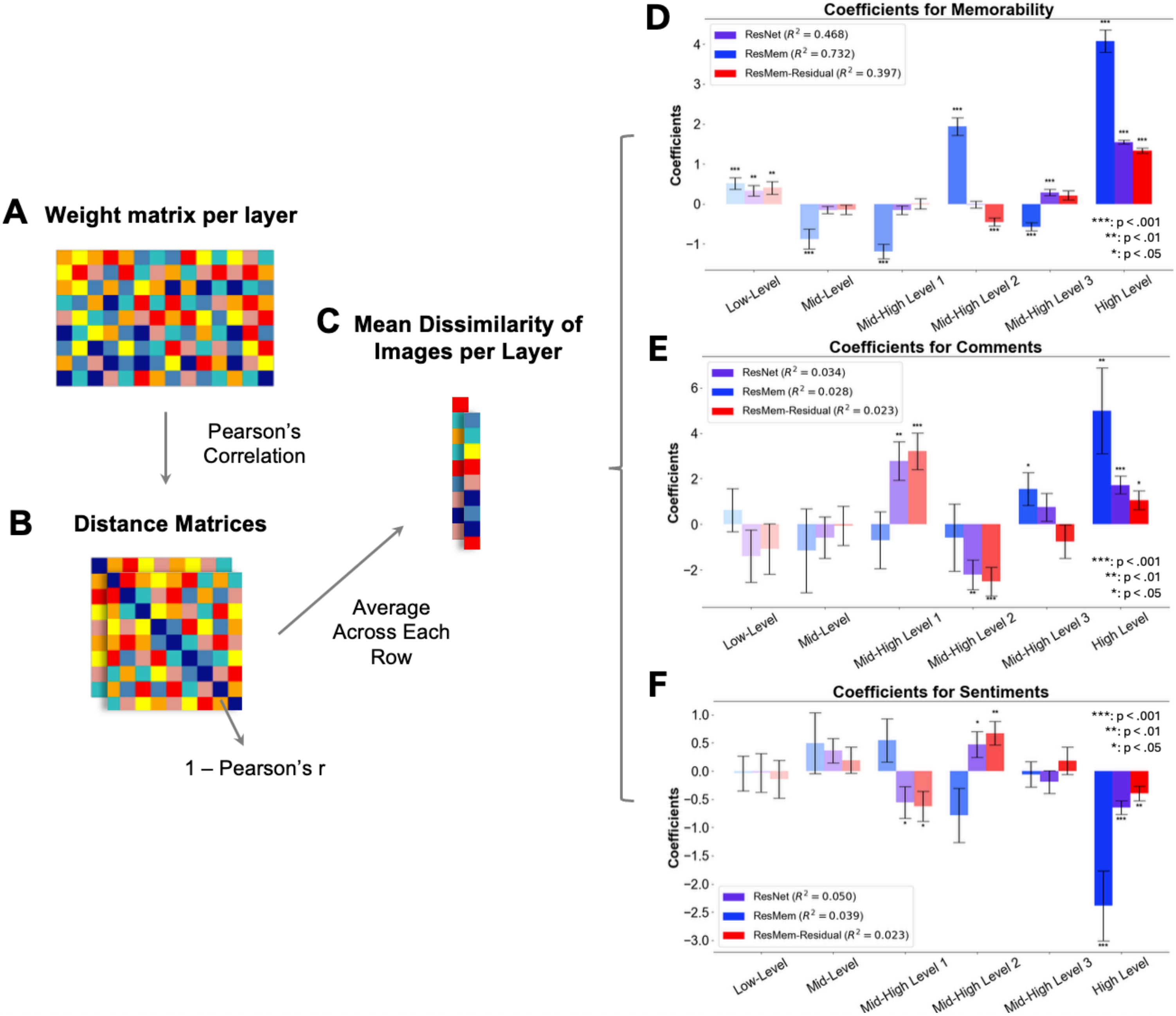

**Fig. 5. Generation of Mean Dissimilarity Scores per Convolutional Layer and Results of Regression Coefficients. (ABC)** We first extracted the outputs of the convolutional layers across all images. A Pearson correlation was calculated with the output of each image and the output of all other images to generate a distance matrix with each cell representing 1 – Pearson's r, indicating the dissimilarity of one image compared with the others as represented in the convolutional layer. This matrix was then averaged across all images to derive a score representing the mean dissimilarity of one image compared with all other images in each layer. This provides a way

to quantify the distinctiveness of different feature representations as reflected in the layers. **(DEF)** Stage underlies bar plots are: Low-level: low-level visual feature (stage 1); Mid-level: mid-level visual features (stage 2); Mid-High level: a) early: abstraction initiation; b) middle: intermediate abstraction; c) abstraction refinement (stage 3); High-level: high level semantic features (stage 4). The bar plots with error bars display the regression coefficients of each stage in the GLMs, where the dependent variables are the memorability scores (**D**), number of comments (**E**), and average sentiment scores (**F**). ResMem-residual is derived from regressing out the shared variances between ResMem and ResNet to derive the memorability-specific contribution by accounting for the image category effect.

## 4. Discussion

We acquired and analyzed social media data to address three core research questions: 1) can memorability predict social media virality; 2) how and why memorability relates to human behavior leading to virality; and 3) what underlying visual features contribute to memorability, virality, and sentiment. To address the first question, we found that memorability as predicted by a neural network (ResMem) is associated with more comments and also partially associated with more upvotes (post scores). To address the second question of "how", we utilized NLP techniques on the comments and discovered that memorability may potentially guide viewers towards external associative thinking, connecting the content to ideas or experiences beyond what is directly shown in the image. This, in turn, may explain "why" memorability attracts more comments, as it may engage individuals in deeper cognitive processing (Kirsh, 2010) through external associative thinking, which thus encourages them to respond (Alhabash et al., 2015). In addition, these comments tend to be less emotionally valanced in nature, demonstrating that memorability's association with virality may be distinctive from the previously found emotion-virality effect (Alhabash et al., 2015, 2019; Berger & Milkman, 2013; Heimbach & Hinz, 2016). In answering the third question, we analyzed the contribution of image representations by examining the convolutional layers of ResMem-residual, in which we found that semantic information is a common underlying predictor for memorability, virality, and sentiment.

**4.1. From Memorability to Virality**. We replicated the correlation analysis between memorability and virality metrics across three different subreddits at three time points to investigate the predictability of memorability on virality. Our results demonstrated that memorable images consistently receive significantly higher numbers of comments while not exactly for upvotes. We did observe a trend of memorable images receiving more upvotes when analyzing the full dataset, but when we removed outliers (which include highly viral posts), the relationship was no longer clear. One possible reason for the lack of an observed effect of upvotes may stem from Reddit's algorithms. Reddit does not display the exact number of upvotes publicly. Instead, it shows a post score calculated using a proprietary algorithm that combines upvotes and downvotes, with additional adjustments to mitigate spam attacks (Robertson, 2016). As such, the interpretation of a lack of upvotes effect does not necessarily indicate such an effect does not exist. Furthermore, one might argue that the titles of the posts could mediate any relationship between memorability and our measures of virality. However, this does not appear to be the case, as humans demonstrate a worse-than-chance ability to predict the virality of posts based solely on their titles (Deza & Parikh, 2015), and we observed that the link between memorability and number of comments remained when controlling for post title. Moreover, certain image categories may inherently have a higher likelihood of going viral or becoming more memorable, as supported by the significant coefficients of ResNet in **Figure 5D,E**. However, the significant coefficients of ResMem-residual indicate that memorability-specific components also capture unique variance in virality. This suggests that the relationship between memorability and virality is not solely driven by image category but instead involves unique aspects of memorability itself.

Previous studies have identified emotional valence and moral content as key stimulus features driving virality (Alhabash et al., 2015, 2019; Berger & Milkman, 2013; Heimbach & Hinz, 2016), possibly due to their ability to capture attention (Brady et al., 2020). However, these factors are likely subject to individual variability in emotional perception, elements such as age, sex, or culture can significantly alter how one perceives the emotional valence or intensity of an image (Neiss et al., 2009; Thompson & Voyter, 2014; Barrett et al., 2011). This variability can thus introduce inconsistency, as evidenced by conflicting findings on whether positivity or negativity is the key virality impacting factor (Alhabash et al., 2019; Berger & Milkman, 2013; Heimbach & Hinz, 2016). At the same time, it is important to acknowledge that certain emotional responses can show broad consistency, particularly toward highly evocative stimuli. Nevertheless, even in these cases, substantial variability often persists in the nuance and intensity of individual reactions, which can complicate their predictive utility. Our study addresses this limitation by introducing image memorability as a more statistically consistent and generalizable predictor of virality. Although derived from human behaviors, memorability functions as a cognitive property that exhibits high agreement across observers. Unlike emotional valence or moral framing which are more context-dependent, memorability reflects image features that tend to influence memory performance similarly across individuals (Isola et al., 2011b), across cultures (Jeong, 2023), age groups (Guo & Bainbridge, 2023), tasks (Bainbridge, 2020), and environments (Davis & Bainbridge, 2023).

Prior work has shown that its impact is relatively robust to variation in prior experience or attentional state, as indicated by minimal interaction between memorability and sustained attention in predicting memory outcomes (Wakeland-Hart et al., 2022; Roberts et al., 2024). Moreover, the automatic calculation of memorability by human observers (Bainbridge et al., 2017), without the need for conscious effort to predict which images are memorable (Isola et al., 2014), suggests that memorability effects arise during bottom-up processing, whereas appraisals of emotion and morality typically involve top-down processes. Additionally, a recent study finding that memorability is correlated with artwork fame further underscores the significance of memorability as a virality-impacting factor (Davis & Bainbridge, 2023).

By analyzing the semantic properties within the comments, we found that memorable images tend to provoke more externally-related thinking rather than thoughts about the contents within the image. This external perception aligns well with prior behavioral studies finding that memorability is a high-level dominated image feature, which presumably conveys semantic meaning underlying the image rather than low-level basic visual features (Bainbridge et al., 2013, 2017). Extending beyond behavioral data, results from electroencephalography (EEG) found that memorable images facilitated high-level perceptual processing and semantic activation (Deng et al., 2024), which relates to processes of high-level thinking as shown in studies on both nonhuman primates and humans (Koechlin et al., 2000; Genovesio et al., 2005; Rodriguez et al., 2023). Further, studies leveraging CNNs have shown semantic information being the primary driver for memorability, as indicated by more contribution from the later convolutional layers, which presumably process high-level semantic information (Kramer et al., 2023; Koch et al., 2020). One fresh finding is that memorability guides people towards external divergent thinking, which to our knowledge, has not been observed by any prior studies. According to the excitation transfer theory (Zillmann, 1971; Cantor & Zillmann, 1973), the arousal elicited by an initial stimulus does not dissipate immediately, but instead likely escalates a subsequent action or emotional states by transferring this residual excitation. In the context of virality, exposure to online content is proposed to create an excitation residual window (Alhabash et al., 2015). This window may drive responses to ease that residual like upvotes, shares, with commenting being particularly likely when users invest in more cognitive resources. Prior research (Kirsh, 2010) has shown that establishing external associations requires more cognitive resources, and memorability may guide such associations and attract more comments. Hence, it is plausible to argue that memorability can trigger excitation arousal, which is also supported by significant memorability-driven activation found in the high-level perceptual regions and medial temporal lobe (Bainbridge & Rissman, 2018). By connecting these findings, we propose that memorability may promote external associative thinking, which requires greater cognitive effort. As a result, users are more likely to engage in commenting, ultimately contributing to virality, a hypothesis that can be investigated in future research.

**4.2. Memorability and Sentiment.** An intriguing while somewhat counterintuitive finding emerged from both the regression and correlation results, indicating a negative association between memorability and the sentiment intensity expressed in post comments. Specifically, images exhibiting higher memorability, and thus greater viral potential, tend to have less emotionally intense discussions. This finding adds nuance to the prevailing understanding in the literature, which has often linked emotional stimuli to increased virality (Alhabash et al., 2015, 2019; Berger & Milkman, 2013). Prior research has consistently shown that emotional content is more likely to elicit emotional responses (Niedenthal et al., 1999), a reaction typically associated with activation in the amygdala, a brain region known to be involved in emotional processing (Hamann & Mao, 2002; Zald, 2003). Likewise, recent findings also showed that viewers' continuous emotional reactions predict social media engagement (Bacic & Gilstrap, 2023). However, our current findings indicate that memorability may serve as a separate, orthogonal predictor of virality, distinct from emotionality. In other words, the characteristics that make content memorable and therefore viral do not necessarily overlap with those that evoke strong emotional reactions.

It is also important to contextualize this distinction within methodological frameworks. Previous studies of virality have predominantly relied on pseudo-experimental conditions, such as lab-based interfaces (Brady et al., 2020), social media simulations in the lab environment (Alhabash et al., 2015, 2019), and text-only contents like lengthy magazine articles (Heimbach & Hinz, 2016). These approaches, while valuable in providing causal inferences, may not fully capture the dynamics of real-world social media environments. Until now, virality in-the-wild on a modern image-based social media platform had only been rarely explored. As such, this methodological difference may suggest that people's emotional responses in a fully naturalistic environment may potentially differ from those observed in a controlled lab-based setup or using platforms organized more around text-based discussion (e.g., X/Twitter). The current findings therefore underscore the importance of investigating virality in authentic, real-world settings, and suggest that memorability-driven virality may unfold through mechanisms that differ fundamentally from traditional emotion-driven models.

**4.3. Underlying Semantic Features from Neural Networks**. By dividing neural network processing into stages, we found that both low-level visual features and high-level semantic features significantly impact memorability estimates, with semantic features serving as the primary driver across all model components. ResNet, as an ImageNet-pretrained neural network, inherently encodes image category effects within its embeddings. When this category effect is regressed out from ResMem, the dominance of semantic features in influencing memorability-specific components remains evident, as demonstrated by the ResMem-residual (**Fig. 5D**). To examine features across different depths, we adopted an exploratory approach: sampling the last convolutional layer in each third of ResNet's residual blocks. This provided a systematic and interpretable way to probe different stages of the network, and future work can build on and refine this strategy. Notably, this finding generally aligns with previous studies showing higher order semantic information contributes most to memorability (Kramer et al., 2023; Koch et al., 2020), supporting this methodology and offering potential guidelines for future research on image feature contributions using deep networks like ResMem.

Our results also lend evidence towards an ongoing debate on whether memorability is more related to prototypicality or distinctiveness of the stimuli (Kramer et al., 2023; Koch et al., 2020; Vokey & Read, 1992; Lee et al., 2000). Although both intermediate and late layers of the ResMem DNN predict memorability, their predictions differ in direction, where prototypicality at an intermediate level but distinctiveness at a late layer are both predictive of memorability. We think that image distinctiveness and prototypicality are not mutually exclusive but instead may coexist and interact. In fact, these results could reflect a growing idea that what makes an image memorable is a delicate balance of prototypicality and distinctiveness, an image must be easy to process (Geotschalackx et al., 2019), and align with our cognitive templates (Deng et al., 2024), yet also have distinctive features (Koch et al., 2020; Lee et al., 2000) to make them stand out. This may also explain why these same mixed patterns occur when predicting number of comments and comment sentiment, which are both tied to memorability. Indeed, a study by Koch and colleagues (2020) using a shallower DNN (AlexNet) observed that memorable images tend to be visually distinctive but semantically prototypical. Further, recent behavioral evidence shows that memorable images tend to be those that are similar to same-category exemplars but distinctive from other-category items (Lee et al., 2024). Hence, we believe this dual pattern in the ResMem DNN reflects this intricate balance of prototypicality and distinctiveness necessary to produce memorable items.

At the same time, an alternate possibility is that this discrepancy could be attributed to differences in the models used. Prior studies examining typicality effects in memorability employed AlexNet or VGG-F, both of which are shallower CNNs with less than 20 convolutional layers (Krizhevskey et al., 2012; Simonyan & Zisserman, 2014), compared to ResMem, which has 152 convolutional layers from ResNet-152 (He et al., 2015). The greater depth of ResMem may enable the extraction of more nuanced or complex features, leading to a shift in how semantic information is weighted in memorability predictions. Additionally, ResMem, a CNN specifically trained to estimate memorability, might capture unique features that optimize memorability predictions, separate from visual classification networks like AlexNet or VGG-F that more closely capture the goals of the human visual system. This idea is supported by findings that ResMem and humans predict different aspects of memorability (Zhao et al., 2024). Furthermore, MemNet (Khosla et al., 2015), the first CNN trained for memorability, is found to learn distinct image patterns or regions that general visual CNNs do not, again supporting the idea that specialized networks can detect different drivers of memorability.

**4.4. Limitations**. Although our study contributes to the literature by identifying a consistent predictor of social media virality, it is not without limitations. Most notably, our analysis focuses exclusively on Reddit, which may restrict the generalizability of our findings to other platforms like Twitter or Facebook. However, Reddit’s structure, particularly its image-centric subreddits, provide a justified context for investigating image-specific virality dynamics. Platforms such as Instagram or YouTube may offer alternative venues, but each poses challenges:

Instagram content is often private and limited to followers, and YouTube centers on video, which falls outside the scope of ResMem, a model tailored to image memorability.

Another limitation is our use of relatively simple vector embeddings (e.g., GloVe) to analyze the relationship between image content and user comments. We did not utilize more advanced language models (LLMs) like BERT (Devlin et al., 2018) or GPT (Radford et al., 2019), which could potentially offer a more nuanced understanding of comment contents. However, we opted for a simpler approach based on the rationale that if straightforward methods yield meaningful results, they offer advantages in interpretability and efficiency. Additionally, our analysis focused on the top five most-liked comments for each post, helping us retain higher-quality, topically relevant content that serves as a proxy for community endorsement. This focus on prominent comments likely enhanced the alignment between comment content and post content, increasing the effectiveness of using simpler methods to explore image–comment similarity. Nonetheless, we acknowledge that future work employing more sophisticated models may reveal additional layers of insights or may extend the current findings by investigating whether there exists a more common or individualized commenting style when encountering memorable images.

**4.5. Practical Relevance**. Given the social media context of this study – a landscape saturated with content creators – the findings offer concrete, data-driven strategies for boosting engagement. By leveraging image memorability scores, easily obtained via the user-friendly ResMem tool (https://brainbridgelab.uchicago.edu/resmem/), creators can design content that is not only visually compelling but also cognitively memorable. This shifts content creation from a trial-and-error process to one grounded in empirical evidence. Furthermore, since memorability is inherently tied to how well content is retained in memory, creators can strategically build on this effect by producing follow-up content that references or reinforces previously shared high-memorability posts. When users have already encoded a viral image into memory, subsequent related content whether thematically, visually, or narratively connected, can tap into that existing cognitive footprint, increasing the likelihood of recognition, engagement, and sharing. This form of temporal content chaining not only deepens audience connection but also amplifies the reach and longevity of the original post's impact.

**4.6. Future directions**. A central issue surrounding social media content is its role in the spread of misinformation. Future work could investigate how memorability may relate to the veracity of social media content and influence the spread of misinformation. Is highly memorable but false information also more likely to be shared, thus amplifying its reach and potential impact?

Moreover, our findings highlight that memorable content tends to be associated with higher immediate engagement. Future research can build on this by assessing whether this memorability-induced engagement translates into long-term effects, in areas like education, where the retention and application of learned materials are critical. Research in aging healthcare could explore how memorability influences information retention in individuals with cognitive impairments (Bainbridge et al., 2019). Further, researchers could explore such potential long-term impacts of memorability in marketing, such as whether more memorable advertisements can lead to real-world outcomes like influencing customer purchasing behavior.

**4.7. Conclusion**. The findings from this study highlight an interesting association between memorability and the spread of content in digital environments, underscoring the potential role memorability may play in driving content dissemination. Notably, the relationship between memorability and virality, plausibly intertwined with external-associative abstract thinking, provides a compelling perspective on how the visual properties of an image can resonate with cognitive processes, influencing human behavior in ways that transcend simple memory recall. This connection suggests that certain semantic elements within an image, as perceived by humans, are associated with the activation of broader networks of related ideas, potentially enhancing both the depth and breadth of the content's impact. These insights underscore the potential for marketers, content creators, or communicators to strategically manipulate the memorability of their contents, thereby fundamentally shaping how information spreads and influences public engagement to amplify its reach and impact.

**Acknowledgements**

This research was supported by a Scialog grant (Research Corporation for Science Advancement) to W.A.B. We thank Yichen Dai for initial piloting of this project idea.

**Author Contributions**

Conceptualization: S.P. and W.A.B. Methodology: S.P and W.A.B. Data collection: S.P. Data Analysis: S.P. Visualization: S.P. Supervision: W.A.B. Manuscript writing and editing: S.P. and W.A.B.

**Data and Code Availability**

The data supporting our findings are available on the Open Science Framework at: https://osf.io/p2bh6/. The codes used to support our findings are available on GitHub at https://github.com/shikangpeng/memMedia.

**Competing Interests**

The authors declare no competing interests.

**Supplementary Information** is available for this paper.

**Correspondence and requests for materials** should be directed to Shikang Peng, Rotman Research Institute at Baycrest; 3560 Bathurst St, Toronto, Ontario, Canada, M6A 1W1; Email: speng@research.baycrest.org.